\documentclass{mn2e}
\usepackage{url} 
\usepackage{amsmath}
\usepackage{amssymb}
\usepackage{graphicx}
\usepackage{epsfig}
\usepackage{ulem}
\usepackage{color}
\usepackage{IEEEtrantools}

\newcommand{\apj}{ApJ}
\newcommand{\apjl}{ApJ}

\newcommand{\aap}{A$\&$A}

\newcommand{\mnras}{MNRAS}

\newcommand{\prd}{PRD}

\newcommand{\nat}{Nature}
\newcommand{\nar}{New Astron. Rev.}
\newcommand{\prc}{Physical Review C}
\newcommand{\physrep}{Physics Reports}

\newcommand{\be}{\begin{equation}}
\newcommand{\ee}{\end{equation}}

\begin{document}
\label{firstpage}

\title[``Spin-flip" and GWs from newly born magnetars]{Neutron star bulk viscosity, ``spin-flip"  and GW emission of newly born magnetars}
\author[]{S. Dall'Osso$^{1,}$\thanks{contact address: sim.dall@gmail.com},   
L. Stella$^2$, C. Palomba$^3$\\
$^1$ Department of Physics \& Astronomy, Stony Brook University, Stony Brook 11579 NY, USA \\
$^2$ INAF - OAR, via di Frascati 33, 00078, Monte Porzio Catone, Roma, Italy \\
$^3$ INFN, Sezione di Roma, P.le A. Moro, 2, I-00185 Roma, Italy}

\maketitle
\label{firstpage}

\begin{abstract}
The viscosity-driven ``spin-flip" instability in newly born magnetars with interior toroidal magnetic fields is re-examined. We calculate the bulk viscosity coefficient ($\zeta$) of cold, $npe \mu$ matter in neutron stars (NS), for selected values of the nuclear symmetry energy and in the regime where $\beta$-equilibration is slower than characteristic oscillation periods. We show that: 
i) $\zeta$ is larger than previously assumed and the instability timescale correspondingly shorter; 
ii) for a magnetically-induced ellipticity $\epsilon_B \lesssim 4 \times 10^{-3}$, typically expected in newborn magnetars, spin-flip occurs for initial spin periods $\lesssim 2-3$ ms, with some dependence on the NS equation of state (EoS). 
 
 \noindent
We then calculate the detectability of GW signals emitted by newborn magnetars subject to ``spin-flip", by accounting also for the reduction in range resulting from realistic signal searches. 
For an optimal range of $\epsilon_B \sim (1 - 5) \times 10^{-3}$, and birth spin period $\lesssim 2$ ms, we estimate an horizon of $\gtrsim 4$ Mpc, and $\gtrsim 30$ Mpc, for Advanced and third generation interferometers at design sensitivity, respectively.
A supernova (or a kilonova) is expected as the electromagnetic counterpart of such GW events.

\noindent
Outside of the optimal range for GW emission, EM torques are more efficient in extracting the NS spin energy, which may power even  
brighter EM transients. 

\end{abstract}


\section{Introduction}
The first detections of GWs from binary black holes (BH; Abbott et al. 2016, Abbott et al. 2017a) and from a binary NS merger (Abbott et al. 2017d) have opened a new era 
in physics and astronomy. Newly born magnetars have long been discussed as a class of compact objects of potential relevance  
for current and future GW detectors (Cutler 2002, Stella et al. 2005, Dall'Osso et al. 2009, Corsi \& Meszaros 2009, 
Dall'Osso et al. 2015). 

The ability of newborn magnetars to emit a distinctive GW signal, the properties of such signals, and their rate of occurrence, 
are sensitive to the NS properties; therefore, they hold the potential to probe the physics of NS interiors. 
Cutler (2002) first pointed out that millisecond spinning NS with predominantly toroidal interior B-fields, e.g. magnetars, may be subject to a secular instability first discussed by Jones (1976), which favors intense GW emission.
Schematically\footnote{see 
Cutler (2002) and Jones (1976) for more details.}, a strong toroidal field deforms the NS shape into a prolate ellipsoid, which undergoes freebody
precession. Dissipation of the precession energy, due to the NS interior viscosity, will drive 
the symmetry (magnetic) axis of the ellipsoid orthogonal to the spin axis, thus maximizing GW emission efficiency. 
This is often referred to as\footnote{Even though, in the observer's frame, it is the magnetic symmetry axis that flips.} 
``spin-flip instability";  the {\it prolate} shape of the ellipsoid, induced by a strong {\it toroidal} magnetic field in the NS core, is essential for the instability. 

The B-field strength in magnetar cores can only be inferred from observations of the galactic population. Based on the energetics of the Dec 27, 2004 Giant Flare 
of the Soft Gamma Repeater SGR 1806-20, Stella et al. (2005) derived a lower limit $\sim 10^{16}$ G for the volume-averaged B-field {\it at birth}. 
Makishima et al. (2014), based on the possible precession of the Anomalous X-ray Pulsar 4U 0142+61, estimated B $\sim 10^{16}$ G at $\sim 10^4$ yrs age. Such magnetic field in the core of newborn magnetars would give rise to GW signals detectable from well beyond the Milky Way.

Key to strong GW emission is that flipping of the symmetry axis be fast compared to other mechanisms like, e.g. magnetic dipole radiation, 
that tap the same energy reservoir as GWs, {\it i.e.} the NS spin. Thus, an effective source of viscosity is crucial.   
To further study this scenario, Dall'Osso et al. (2009) considered bulk viscosity in pure $npe$ NS matter soon after birth, at temperatures 
$\sim 10^{10}$ K, when the NS crust has not yet formed.

These authors concluded that:   
i) for birth spin $\sim$ 1-3 ms, the instability is sufficiently fast if the interior B-field is\footnote{From here on, $Q_n \equiv Q/10^n$.} B$_{{\rm int},16} \lesssim 
4$ and the exterior dipole B$_{\rm d, 14} \lesssim 5$; ii) GW signals are detectable with Advanced LIGO/Virgo from Virgo cluster distances. Given an estimated magnetar formation rate $\sim$ 1 yr$^{-1}$ within that volume, this may lead to an interesting rate of detectable events, in particular for $B_{{\rm int},16} \gtrsim 1$, $B_{{\rm dip}, 14} < 3$, and birth spin periods $< 2.5$ ms.

Here we improve on previous work in several ways: in Sec. \ref{sec:bulk} we calculate the bulk viscosity coefficient of $npe\mu$ matter, for three representative choices of the NS EoS, showing that it is generally larger than previously assumed. In passing, we address doubts raised about the effectiveness of bulk viscosity. In Sec. \ref{sec:spinflip} we summarize the formalism used to model spin-flip as a consequence of viscous dissipation of freebody precession, 
and calculate numerically the time evolution of the tilt angle of the magnetic axis,  along with the corresponding GW luminosity. In Sec. \ref{sec:transients}, using our new results, we re-asses the detectability of newborn magnetars with current and future GW detectors, and further comment on  perspectives for the detection of associated EM signals. 

\section{Bulk viscosity in the NS core}
\label{sec:bulk}
Fluid bulk viscosity is due to pressure/density variations from equilibrium. In a precessing NS, such fluctuations are excited 
at the precession frequency. In $\beta$-stable NS matter, 
pressure depends on the local density {\it and} charged particle fraction: when a fluid element is displaced from
equilibrium, the ensuing compression
will activate $\beta$-reactions, to establish a new
pressure and chemical equilibrium. Bulk viscosity is thus characterised by two timescales: the 
perturbation period, $T_p = 2 \pi / \omega$, equal to the precession period, and the relaxation timescale $\tau_{\beta} = 2 \pi / \omega_{\beta}$, on 
which chemical equilibrium is restored. For $npe$ matter, the bulk viscosity coefficient is\footnote{Eq. \ref{eq:def-bulk-coeff}  is valid for any fluid with relativistic components. For practical purposes, in the case of $npe \mu$ matter, we will adopt a slightly different expression which is derived from (\ref{eq:def-bulk-coeff}).}   (Lindblom \& Owen 2002)
\begin{equation}
\label{eq:def-bulk-coeff}
{\rm Re}(\zeta) \equiv \frac{ n_b \tau_{\beta} \displaystyle \left(\frac{\partial P} {\partial x}\right)_{n_b} \frac{dx}{dn}}{1+ (\omega \tau_{\beta})^2}  = \displaystyle  \frac{z}{1+z^2} \frac{n_b}{\omega}\left(\frac{\partial P}{\partial x}\right)_{n_b} \frac{dx}{dn_b} \, .
\end{equation}
where $P, n_b$ are the total pressure and baryon density, $z = \omega \tau_{\beta}$ and $x=n_p/n_b$ the proton fraction. 
Eq. (\ref{eq:def-bulk-coeff}) highlights the dependence of bulk viscosity on the perturbation frequency, chemical composition, density and pressure profiles of the NS structure. 
~\\

\noindent
{\bf The different regimes --} Eq. (\ref{eq:def-bulk-coeff}) has two main regimes of $\zeta$ as a function of $z$:

i) $z \ll 1$, ``low frequency'' limit. Chemical equilibrium is established quickly compared to the perturbation period ($T_p$). Thus, deviations from chemical equilibrium cannot grow much and energy losses remain very limited, resulting in a small bulk viscosity coefficient, which scales like $\zeta \propto z$; 

ii) $z \gg 1$, ``high frequency" limit. Chemical imbalance is erased over a time much longer than the perturbation period.  
During each cycle, deviations from chemical equilibrium grow {\it almost} freely: the small dissipation due to  
$\beta$-reactions only builds up in a large number of cycles, eventually damping the perturbation. 
In this regime $\zeta \propto z^{-1}$.

These two regimes join smoothly around $z \sim 1$, where the bulk viscosity coefficient $\zeta(z)$ reaches a maximum.
~\\

\noindent
{\bf Standard expression --}  $z \gg 1$ is typically the relevant regime in NS (e.g. Haensel, Levenfish \& Yakovlev 2000, 2001, Lindblom \& Owen 2002, 
Dall'Osso et al. 2009), 
unless $T > 10^{10}$ K. 
In particular, assuming a NS 
made of pure $npe$ matter, and treating each particle species as a fluid of non-interacting, fully  
degenerate fermions, the standard expression for the bulk viscosity coefficient (Sawyer 1989) can be derived from 
Eq. \ref{eq:def-bulk-coeff} for $z \gg 1$ 
 \be
 \label{eq:zeta-stand}
 \zeta^{({\rm std})} \approx 6 \times 10^{-59} 
 \rho^2 T^{6}  \omega^{-2} \, .
 \ee
Eq. \ref{eq:zeta-stand} can be improved in two ways: {\it a)} a more realistic description of NS matter, which accounts for the   
interactions among baryons, by specifying the NS EoS;  
{\it b)} the inclusion of additional particles, expected to appear in the NS core at large densities (e.g., Haensel et al. 2000, Lindblom \& Owen 2002). Muons will be first produced in $\beta$-reactions once the electron Fermi energy exceeds the muon rest-mass $\approx 105$ MeV. The exact density threshold for muon production depends on the NS EoS and has a typical value $\rho \lesssim  2.3 \times 10^{14}$ g cm$^{-3}$.  
We will not consider further particles, that might appear in the core at $\rho > 8 \times 10^{14}$ g cm$^{-3}$.

\subsection{Nuclear symmetry energy}
\label{sec:symmetry}
Baryon interactions in the NS EoS are described in terms of $E_N(n_b, x)$, the nucleon energy per baryon, at baryon number density $n_b = n_n+n_p$.
If E$_N(n_b, 0)$ is the energy of pure neutron matter and $E_N (n_b, 1/2)$ the energy of symmetric matter, 
then the former exceeds the latter by the symmetry energy, $S_0(n_b)$. For intermediate $x$-values, the excess energy is obtained interpolating between these two limits
\be
\label{eq:def:symmetry}
{\rm E}_N(n_b, x) \approx {\rm E}_N(n_b, 1/2) + S_0 (n_b) (1-2x)^2 \, .
\ee
$S_0(n_b)$ has a kinetic and a potential energy component
\be
\label{eq:symmetry}
S_0 (u) = S_k u^{2/3} + S_v u^\gamma \, ,
\ee
the latter incorporating baryons interactions. In Eq. \ref{eq:symmetry}, $u = n_b/n_s$, $n_s\approx 0.16$ fm$^{-1}$ is the baryon number density at the nuclear saturation density $\rho_s \approx 2.7 \times 10^{14}$ g cm$^{-3}$, $S_k = 17$ MeV and $S_k+S_v = S_0(u_s)$. The index $\gamma \sim 0.2-1$ (Steiner et al. 2010) parametrizes the uncertain scaling of the potential energy with density. Eq. \ref{eq:symmetry} is often written as $S_0(n_b) = S_{\nu} \left(n_b/n_s\right)^{\Gamma}$, where $S_{\nu} = S_k+S_v \sim 30-34$ MeV positively correlates with $\Gamma\sim 0.45-0.7$ (Lattimer \& Prakash 2016). Here, we consider three representative choices of $(S_v, \gamma)$ that span the range of uncertainty on both parameters: chosen values are reported in Tab. \ref{tab:xnpe}. Case $I$  gives a relatively stiff EoS approximating the  
\begin{figure}
\begin{center}
\includegraphics[scale=0.295]{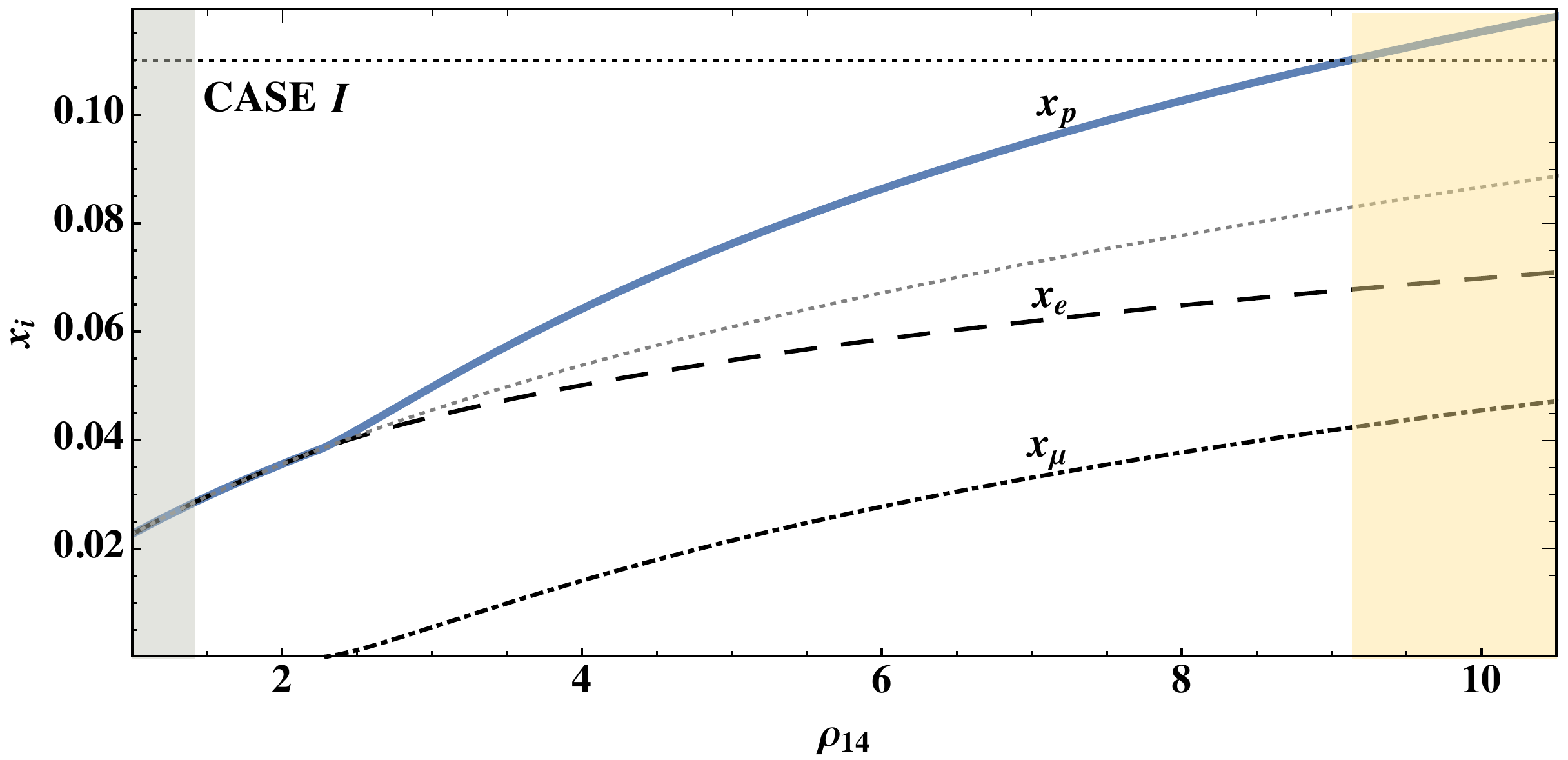}
\includegraphics[scale=0.295]{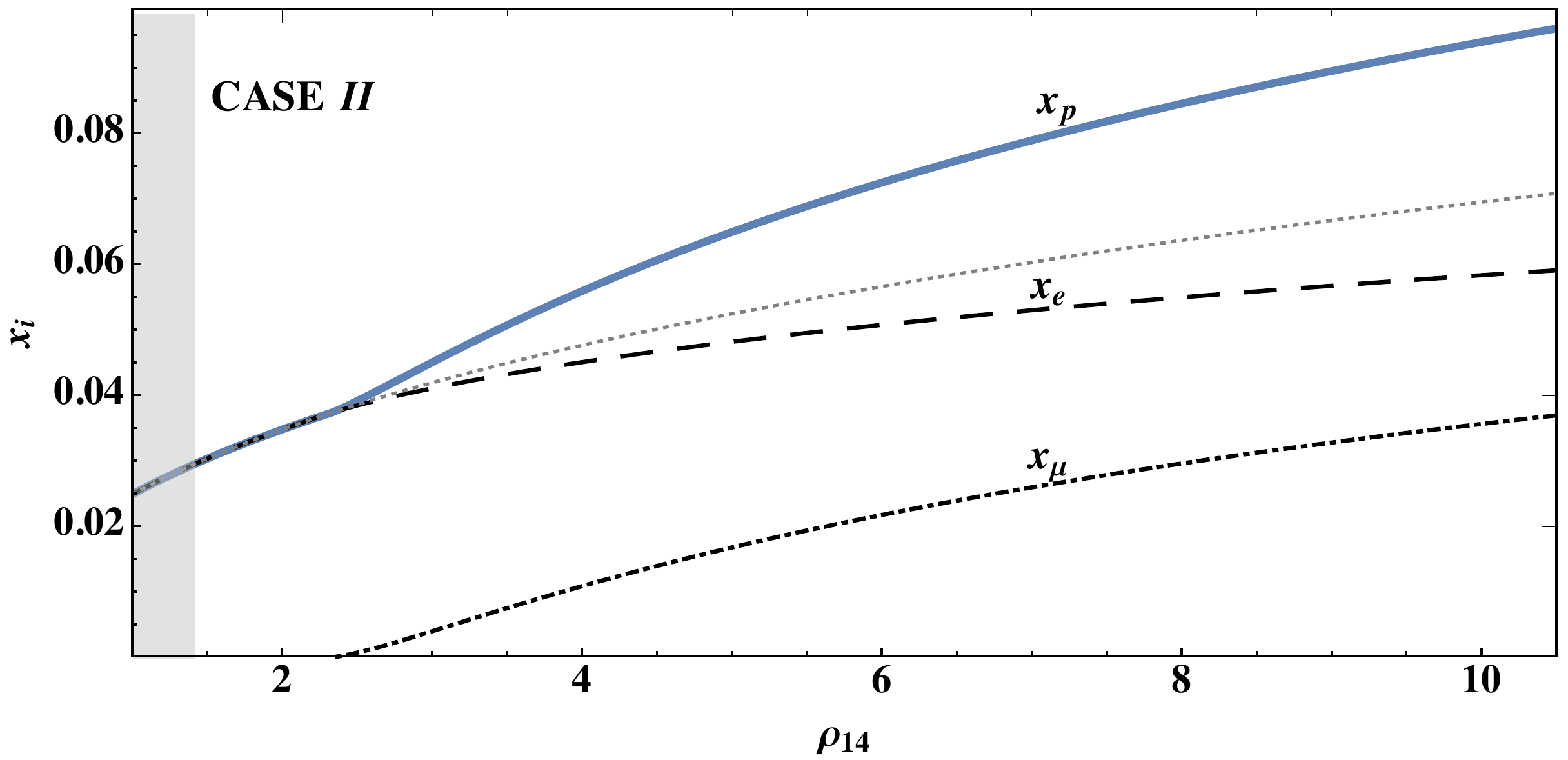}
\includegraphics[scale=0.295]{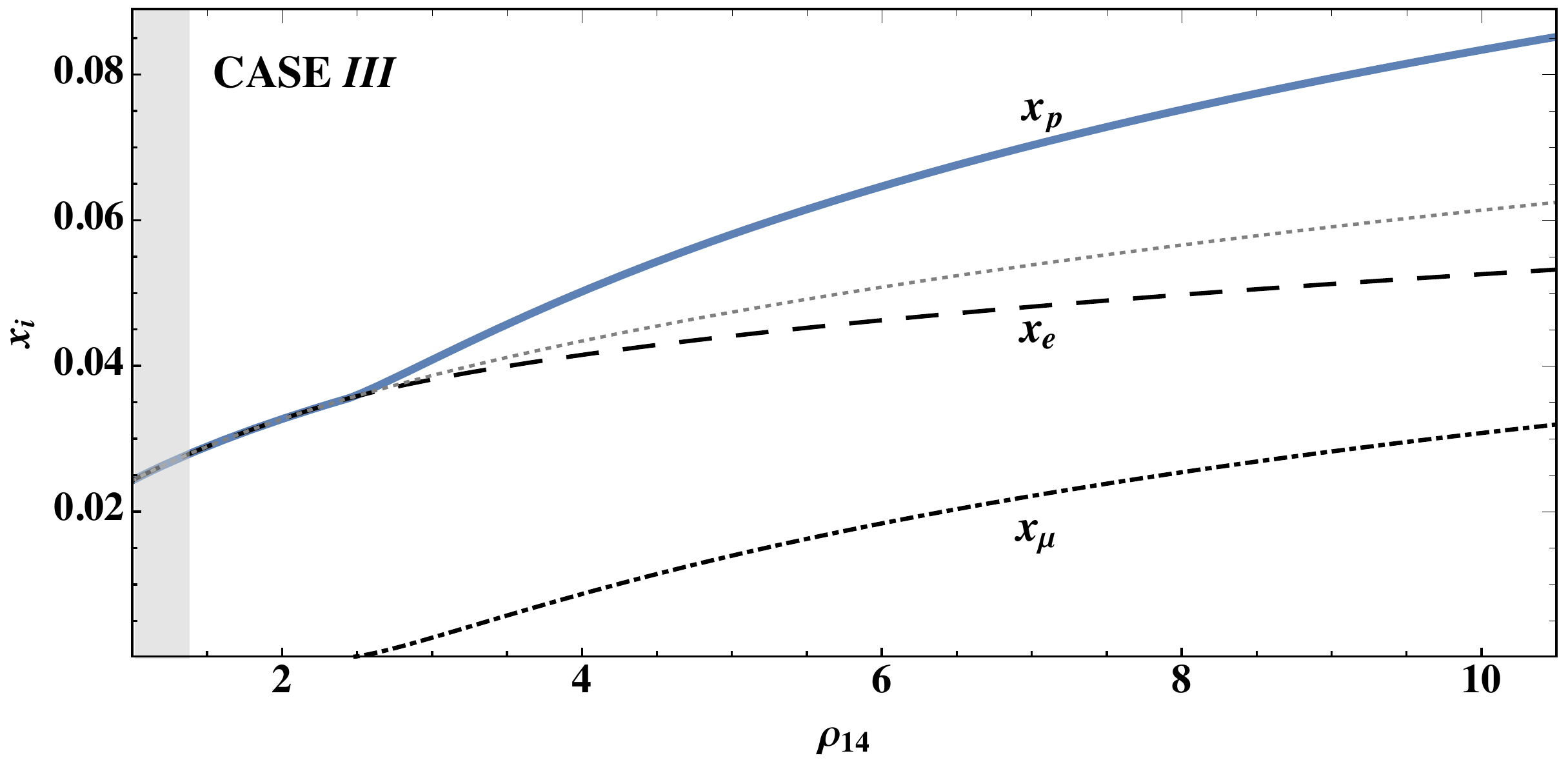}
\caption{Proton, electron and muon fractions for the three cases in Tab. \ref{tab:xnpe}. {\it Upper Panel}: the horizontal dotted line is the $x_p$ threshold for direct Urca reactions. It is not reached in cases $II$ and $III$. In the yellow area direct Urca reactions are allowed, hence our calculations should be modified. The grey area on the left corresponds to densities of the NS crust.}
\label{fig:protonfractions}
\end{center}
\end{figure}
APR EoS (Akmal, Pandharipande \& Ravenhall 1998). Cases $II$ and $III$ give a progressively softer EoS, consistent with a NS maximum mass $\gtrsim 2 M_\odot$ (e.g. Demorest et al. 2010).
\subsection{Charged particle fraction} Accounting for baryon interactions increases the charged particle fraction with respect to the simple case with no interactions.
~\\

\noindent
{\bf Case 1. $npe$ matter -- } the total energy per baryon inside a NS includes,  besides E$_N$, the energy of relativistic electrons, E$_e = 3/4 x E_{F_e} = 
(3/4) \hbar c x (3 \pi^2 n_b x)^{1/3}$, with $x = x_e = x_p$ due to charge neutrality. The equilibrium composition is obtained by minimizing the total energy with respect to $x$,  $\partial \left[E_N (n_b, x) + E_e(n_b, x)\right]/\partial x =0$, which is equivalent to imposing the equality of the chemical potentials, $\mu_e = \mu_n - \mu_p$
\begin{equation}
\label{eq:x}
\hbar c \left(3 \pi^2 n_b x \right)^{1/3} = - \frac{\partial E_N}{\partial x} = 4 S_{\nu} \displaystyle \left(\frac{n_b}{n_s}\right)^{\Gamma} (1-2x)  \, .
\end{equation}
\begin{table}
\caption{Different choices for the symmetry energy $S_\nu$ and power-law index $\gamma$. Case $I$ matches closely
the results for the APR EoS. Cases $II$ and $III$ are for illustration purposes.}
\begin{center}
\begin{tabular}{c|c|c|c|c}
 CASE & $S_v$[MeV] & $\gamma$ & $S_\nu$[MeV] & $\Gamma$ \\
\hline
 $I$ & 15.5 & 1/2 & 32.5 & $ 0.59 $ \\
 $II$ & 14.5 &  1/3 & 31.5 & 0.525 \\
 $III$ & 13.5 & 1/4 & 30.5 & 0.503 \\ 
\end{tabular}
\end{center}
\label{tab:xnpe}
\end{table}

Solving Eq. \ref{eq:x} gives the proton fraction in the NS core: for example, 
at the nuclear saturation density, $x (n_s) \approx 0.04$ for all EoS considered, as opposed to $x(n_s) \approx 0.006$ obtained in the non-interacting case.
~\\
~\\

\noindent
{\bf Case 2. $npe \mu$ matter -- }: the appearence of muons introduces new constraints. First,  
the electron and muon chemical potentials must be equal. Second, $x_e$ and $x_p$ will appear as different variables in Eq. \ref{eq:x},  
since now $x_p$ must equal the sum of the electron and muon fractions. The 
$e, \mu$ and $p$ fractions are determined from
\begin{eqnarray}
\label{eq:xmuons}
\mu_e & = & \mu_n - \mu_p  \nonumber \\
\mu_\mu & = & \mu_e \nonumber \\
x_p  & = & x_e + x_\mu  \, .
 \end{eqnarray}

The profiles $x_p(n_b), x_e(n_b)$ and $x_{\mu}(n_b)$, corresponding to the three EoS's of Tab. \ref{tab:xnpe} are shown Fig. \ref{fig:protonfractions}. The threshold for muons production is $\rho_\mu \approx 2.5 \times 10^{14}$ g cm$^{-3}$:  below $\rho_\mu$, $x_\mu =0$ while $x_p = x_e$ is determined by  Eq. \ref{eq:x}. Where muons appear, the proton fraction is increased. Accordingly, the threshold for the onset of direct URCA reactions is reached at a somewhat lower density ($\gtrsim 9 \times 10^{14}$ g cm$^{-3}$).

\subsection{Relaxation timescale}
\label{sec:relaxtime}
The relaxation timescale, in the same approximation that gives Eq.~\ref{eq:zeta-stand}, is (Reisenegger \& Goldreich 1992) 
\be
\label{eq:taubeta-standard}
\tau^{({\rm old})}_\beta = \frac{3 n_c}{\lambda E_{F_n}} \approx 6.9 ~{\rm s}~\left(\rho/ \rho_{\rm n}\right)^{2/3} T^{-6}_{10}\, , 
\ee
where $\lambda$ is related to the emissivity of modified Urca reactions and E$_{F_n}$ is the neutron Fermi energy. 
Including $S(n_b, x)$ decreases $\tau_\beta$, which in turn increases $\zeta$ compared to Eq. \ref{eq:zeta-stand}. In order to derive the 
relaxation timescale in this case  
let us assume e.g., a 
density perturbation in 
a fluid element, which will thus 
 find itself out of chemical equilibrium by the amount $\delta \mu =  \mu_n - \mu_p - \mu_e$. This is related to $\delta n_c$, the deviation of the charged particle density from its equilibrium value (Eq.~\ref{eq:x}). $\beta$-reactions will be activated, in order to bring $n_c$ to its new equilibrium value and restore chemical equilibrium ($\delta \mu =0$).
The relaxation timescale is, as usual, 
$\tau_\beta = \delta n_c/\delta \Gamma$, where $n_c = n_p = n_e$, and $\delta \Gamma = \lambda \delta \mu$ is\footnote{Under the assumption that $\delta \mu \ll kT$.} the difference in the rates between direct and inverse $\beta$-reactions. Perturbing Eq. \ref{eq:x} with respect to $x$, we get \begin{equation}
\label{eq:tau}
\tau_\beta \equiv \frac{ 3 n_c}{\lambda \left[E_{F_n} + 24 S_{\nu}~ x (n_b/n_s)^{\gamma}\right]} \, . 
\end{equation}
which generalizes Eq. (\ref{eq:taubeta-standard}).  The symmetry energy term in the denominator is typically $\gtrsim$ 0.5 E$_{{\rm F}_n}$, which reduces $\tau_\beta$ in $npe$ by a factor 1.5 compared to $\tau_\beta^{{\rm (old)}}$. 

\noindent
Allowing for the presence of muons, and accounting for the contribution of the proton branch of different $\beta$-reactions, provides another factor $\sim$ 2 reduction (Haensel et al. 2001). We thus conclude that $\tau_\beta \simeq 1/3 \tau_\beta^{{\rm (old)}}$ for $npe \mu$ matter.
\subsection{Total energy and pressure profile}
The total energy per baryon, for $npe \mu$ matter, is E$_T (n_b, x_p) = E_N(n_b, x_p) + 3/4 x_e E_e(n_b,x_p) + 3/4 x_\mu E_\mu (n_b, x_p)$, where $E_\mu$ is the equivalent of $E_e$.
The total pressure is $P(n_b, x_p) = P_N (n_b, x_p) +P_e +P_\mu$, where $P_e, P_\mu$ are the partial pressures of the free lepton gases.   
The nucleon pressure is defined as 
\be
\label{eq:pressure-def}
P_N (n_b, x_p) = n_b^2 \frac{\partial E_{\rm N} (n_b, x_p)}{\partial x} \, . 
\ee
$p = n^2 \partial \left[E(n, x)+E_e(n,x)\right]/ \partial x$.  
Summing $P_e$ and $P_\mu$ to Eq. \ref{eq:pressure-def} gives the total pressure as a function on $n_b$ and $x$, needed to calculate $(\partial P/\partial x)_n$ in Eq. \ref{eq:def-bulk-coeff}.
\subsection{Bulk viscosity coefficient of $n p e \mu$ matter}
\label{sec:bulk-coeff}
When several particles species are present, the formulation by Haensel et al. (2001) turns out to be more practical to calculate $\zeta$ than that of Eq. \ref{eq:def-bulk-coeff}. We write the total bulk viscosity as the sum of partial bulk viscosities\footnote{For $\tau_\beta > T$.} due to each of the channels for $\beta$-reactions. Thus,  $\zeta = \zeta_{ne}+\zeta_{pe}+\zeta_{n \mu} + \zeta_{p \mu}$, where 
\be
\zeta_{Nl} =  \frac{\lambda_{N l}}{\omega^2} \left| \frac{\partial P}{\partial x} \right |_{n_b} \frac{dx}{d n_b} = \frac{\left| \lambda_{Nl} \right|}{\omega^2} C_{l}^2 \,  ,
\ee
$N l$ standing for each nucleon/lepton couple. The term $C_{l} \equiv n_b \partial \eta_l / \partial n_b$, where $\eta_l = \mu_n - \mu_p - \mu_l$ is the chemical potential imbalance of leptons, and the $\lambda$'s for each branch of modified URCA reactions are given by Haensel et al. (2001). The neutron/proton chemical potentials are $\mu_{n,p} = \partial \left(n_b E_N \right) / \partial n_{n,p}$. For the leptons, $\mu_l = \left(m_l^2 c^4 + p_{F l}^2 c^2 \right)^{1/2}$, where $p_{F l} = \hbar \left(3 \pi^2 n_l\right)^{1/3}$ is the Fermi momentum.

\begin{figure}
\begin{center}
\includegraphics[scale=0.9]{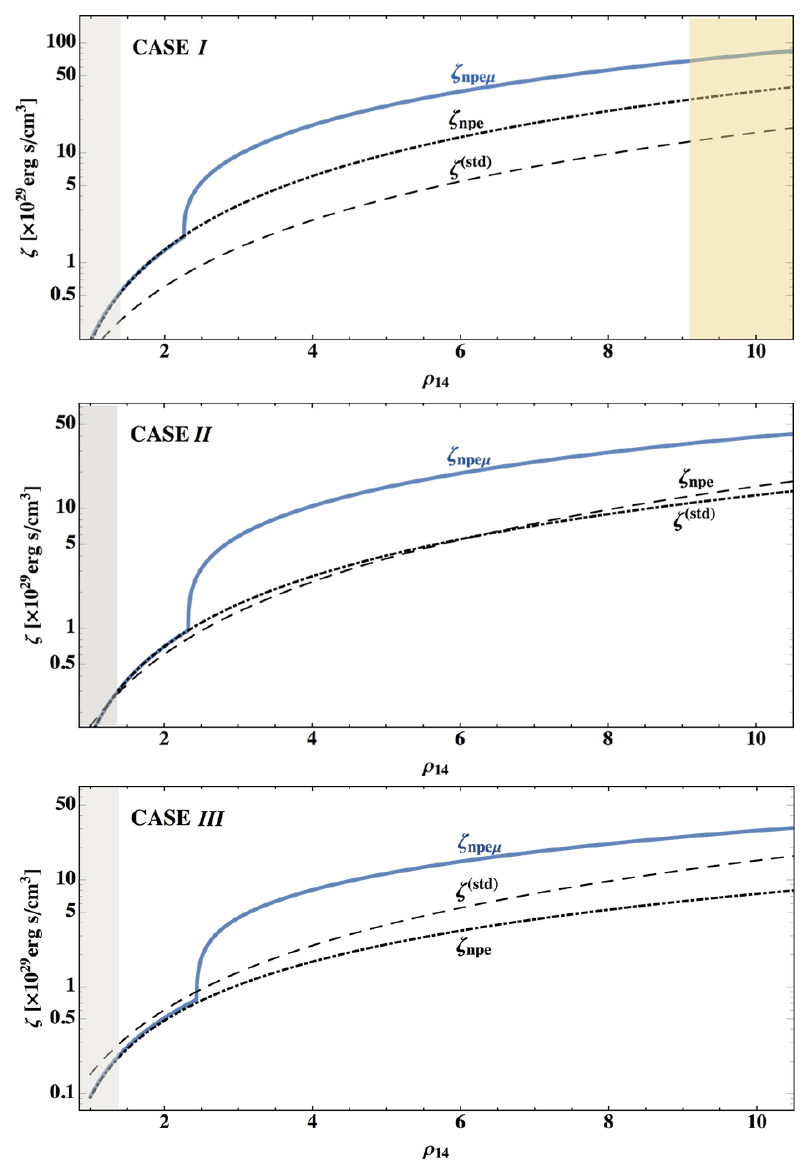}
\caption{Bulk viscosity coefficient for $npe \mu$ NS matter (blue), for the three cases of Sec. \ref{sec:symmetry}. The curves marked $\zeta_{npe}$ represent the coefficient calculated without muons,  those marked $\zeta^{({\rm std})}$ are from the ``standard" expression (Eq. \ref{eq:zeta-stand}). The grey and yellow shaded areas are as in Fig. \ref{fig:protonfractions}. From top to bottom, the density-averaged value of $\zeta$ is about 5, 3 and 2 times larger than that derived from Eq. \ref{eq:zeta-stand}. }
\label{fig:bulk-coefficients}
\end{center}
\end{figure}

\noindent
With these definitions, $C_l$ becomes (Haensel et al. 2000)
\begin{eqnarray}
\label{eq:Cl}
C_l & = & -n_b \frac{\partial^2 E_N(n_b, x_p)}{\partial n_b \partial x_p} - \frac{ c^2 p^2_{F l}}{3 \mu_l} =  \nonumber \\
 & = & \left(1- 2 x_p \right) n_b \frac{ d}{d n_b} \left(\frac{\mu_l}{1 - 2 x_p}\right) - \frac{ c^2 p^2_{F l}}{3 \mu_l}  \, ,
 \end{eqnarray}
where Eq. \ref{eq:symmetry} has been used in the last step. We adopted Eq. \ref{eq:Cl} to calculate the bulk viscosity coefficient of $npe \mu$ matter, for the three EoS's discussed in Sec. \ref{sec:symmetry}. The results are shown in Fig. \ref{fig:bulk-coefficients}: for each case, the blue curve shows the run of $\zeta$ with density, compared with  its value in the absence of muons, and with the value given by Eq. \ref{eq:zeta-stand}. More details are given in the caption.

In all three cases, the resulting bulk viscosity coefficient is greater than the standard one at all densities: their ratio decreases slightly 
with density for EoS's $II$ and $III$, while it is constant for EoS $I$ (above the threshold for the appearence of muons). Assuming a NS with a 12 km radius and 1.4 M$_\odot$ mass, the density-averaged values of $\zeta$ for the three EoS's considered here are, respectively, $\sim 5, 3$ and 2 times larger than the density-averaged value of Eq. \ref{eq:zeta-stand}. These ratios represent a convenient parametrization of the effective increase of NS bulk viscosity, for the EoS's and chemical compositions considered here, and will be adopted in the following.

\section{``Spin-flip"}
\label{sec:spinflip}
We can now calculate the time evolution of the tilt angle in newborn magnetars. Most details were worked out by Dall'Osso et al. (2009), based on  Mestel \& Takhar (1972; see also Lasky \& Glampedakis 2016). 
~\\

\noindent
{\bf NS cooling --} 
We consider modified Urca cooling, that is simply described as (e.g. Page et al. 2006)
\be
\label{eq:nu-cooling}
T_{10} (t) = \left( \frac{t}{20~{\rm s}} +\frac{1}{T_{i,10}^6} \right)^{-1/6} \, ,
\ee
$T_{10} = T/10^{10}$K being the NS temperature and assuming T$_{i, 10}  \gtrsim 3$. Eq. \ref{eq:nu-cooling} describes hot NS where neither protons nor neutrons are superfluid. Protons become superconducting below T$_{cp,10} \sim 5$, or $t \approx 10^3$ s, while neutrons likely become superfluid at much lower T (and later times; e.g. Page et al. 2011). The transition to proton superconductivity reduces progressively the neutrino emissivity, as $T$ drops below T$_{cp}$: however, (\ref{eq:nu-cooling}) remains approximately valid as long as $T/ T_{cp} \gtrsim 0.7$ (cf. Haensel et al. 2001), {\it i.e.} up $t \sim 10^4$ s. 
~\\

\noindent
{\bf Magnetically-induced ellipticity --} As a first approximation, the NS magnetically-induced ellipticity is of the order of the ratio of magnetic to gravitational binding energy, 
\be
\label{eq:epsilonB-rough}
\epsilon_B \sim 4 \times 10^{-4} B^2_{t, 16} R_{12}^4/M^2_{1.4} \, ,
\ee
where $B_{t, 16}$ is the volume-averaged {\it toroidal} field strength in units of $10^{16}$ G, $R_{12}$ the NS radius in units of 12 km and $M_{1.4}$ the NS mass, in units of 1.4 $M_\odot$. Corrections due to the magnetic field geometry in the NS interior can lead to substantially larger deformations at a given B-field strength (Bonazzola \& Gourgoulhon 1996, Mastrano et al. 2011, Akg\"{u}n et al. 2013, Dall'Osso et al. 2015). In particular, the toroidal-to-poloidal field ratio is an unknown parameter that can be very large in the core of non-barotropic NS (Braithwaite 2009, Akg\"{u}n et al. 2013,  Ciolfi \& Rezzolla 2013). 
Given these uncertainties, we will adopt $\epsilon_B \approx 10^{-3}$ as a reference value\footnote{For example, the twisted-torus used in Dall'Osso et al. 2015 has $\epsilon_B \approx 0.9 \times 10^{-3} B^2_{t, 16}$, for the same mass and radius used here.} of $\epsilon_B$ for a $\sim 10^{16}$ G toroidal B-field in the NS core.  
Calling $\chi$ the tilt angle of the magnetic symmetry axis to the spin axis, the freebody precession frequency will be $\omega = \Omega \epsilon_B ~{\rm cos} \chi$, with $\Omega$ the NS spin frequency.
\subsection{Energy dissipation} 
\label{sec:energy-diss}
The energy dissipation rate due to bulk viscosity is 
\be
\label{eq:diss-rate}
\dot{{\rm E}}_{\rm diss}\equiv  \int \zeta |\nabla \cdot \mathbf{\delta v} |^2 \sim \omega^2 \int \zeta \left(\Delta \rho/\rho\right)^2 dV \, ,
\ee
where $\Delta \rho = \delta \rho + \pmb{\xi}\cdot\nabla\rho = -\rho_0 \nabla\cdot\pmb{\xi}$ is the Lagrangian density perturbation due to the precessional motion, $\delta \rho$ the corresponding Eulerian perturbation, $\pmb{\xi}$ the displacement field and $\delta\pmb{v} = \partial_t\pmb{\xi}$ the velocity perturbation.  
~\\

\noindent
{\bf Fluid compression --} Dall'Osso et al. (2009) approximated $\Delta \rho \approx \delta \rho$, adopting $\delta \rho$ as derived by Mestel \& Takhar (1972). Lasky \& Glampedakis (2016) argued that this approximation only holds if $\omega \tau_\beta <1$ whereas, if $\omega \tau_\beta > 1$, $\beta$-reactions would have no time to occur in one precession cycle,  leaving the charged particle fraction unchanged, $\Delta x =0$; since, to first order in the perturbation, the fluid compression $\nabla\cdot\pmb{\xi} \propto \Delta \rho \propto \Delta x$, this\footnote{This proportionality holds strictly to first-order (Mestel \& Takhar 1972). Studying higher-order terms, Lander \& Jones (2017) found that fluid compressibility remains high even as $\Delta x$ is reduced. Our estimates may thus be regarded as conservative.} would give $\Delta \rho =0$, thus quenching bulk viscosity and preventing spin-flip. 

\noindent
By imposing $\omega \tau_\beta \leq 1$, and using the spin-flip timescale (Eq. \ref{eq:dimensional}), Lasky \& Glampedakis (2016) derived a maximum ellipticity $\epsilon_{\rm sf} \approx 5 \times 10^{-3} \rho_{15}/P_{\rm ms}^2$ for spin-flip to operate (P$_{\rm ms}$ is the spin period in milliseconds). Repeating their argument with our updated values $\tau_\beta= 1/3 \tau_\beta^{{\rm (old)}}$ and $\zeta = A \zeta^{{\rm (std)}}$ (Sec. \ref{sec:symmetry}  and \ref{sec:bulk-coeff}), we find $\epsilon^{({\rm new})}_{\rm sf} = A \epsilon_{\rm sf} \approx 1.5 \times 10^{-2} ($A/3)$ \rho_{15}/P_{\rm ms}^2$. The mechanism can thus operate on a wider range of ellipticities than previously suggested: constraints on $\epsilon_B$ based on short GRB observations should be accordingly revised.

The above argument is still approximate, since it (i) uses the simple scaling (\ref{eq:dimensional}), (ii) has an explicit dependence on $\rho$ and (iii) assumes a sharp cut-off of bulk viscosity at $\omega \tau_\beta =1$. To improve on these points we use the definition of the dissipation timescale (Eq. \ref{eq:def:taudiss}), integrating it over the NS density profile, and introduce a slight modification in the treatment of fluid compressibility. 
 First note that $\tau_\beta$ defines the characteristic timescale for $\beta$-reactions in a perturbed fluid element. Because 
a large number of such reactions per unit volume ($\sim \delta n_c$) must occur in the time $\tau_\beta$, a fraction $\sim t/\tau_\beta$ of those reactions must occur in a time interval $t < \tau_\beta$, causing some energy dissipation. 
Consider now a perturbation with the period $T_p < \tau_\beta$. The charged particle fraction will change by an amount that is $\sim T_p/\tau_\beta$ times smaller than when $T_p > \tau_\beta$, and 
 we expect $\Delta \rho$ to be $\sim T_p/\tau_\beta$ smaller than in the long timescale regime, where it was $\approx \delta \rho$. Thus, when $\tau_\beta > T_p$ we assume the relation $\Delta \rho \approx \delta \rho (T_p/\tau_\beta)$: the decreasing efficiency of $\beta$-reactions provides a force that opposes compression but, as long as $T_p \approx \tau_\beta$, cannot prevent it altogether.

\noindent
To summarize, depending on the bulk viscosity regime we will write:  {\it a)} $\Delta \rho \approx \delta \rho$ ($\omega \tau_\beta \leq 1$, highly compressible fluid); {\it b)} $\Delta \rho = \delta \rho (T_p/\tau_\beta)$  in the opposite limit ($\omega \tau_\beta >1$). Note that, because the integral in (\ref{eq:diss-rate}) contains the square of $\Delta \rho$, energy dissipation becomes quickly negligible as $\tau_\beta > T_p$. Thus, our expression models the onset of fluid incompressibility as a smooth, yet fast transition that occurs, as the NS cools, in a narrow region around $\omega \tau_\beta \gtrsim$ 1.

\subsection{Dissipation timescale} 
\label{sec:timescale}

The energy dissipation timescale is (Ipser \& Lindblom 1991)
\be
\label{eq:def:taudiss}
\tau_{\rm diss}   \equiv   \frac{2 E_{\rm pre}}{\dot{E}_{\rm diss}} = \displaystyle \frac{ I}{\epsilon_B} \frac{1}{\displaystyle \int \zeta \displaystyle  \left(\frac{\Delta \rho}{\rho}\right)^2~dV} \, 
\ee
where, to first order in $\epsilon_B$, the  freebody precession energy is E$_{\rm pre} = 1/2 I \Omega^2 \epsilon_B {\rm cos}^2 \chi = 1/2 I \omega^2 \epsilon_B^{-1}$ (Dall'Osso et al. 2009). A simple dimensional analysis of (\ref{eq:def:taudiss}) gives the scaling of $\tau_{\rm diss}$ with the NS parameters 
\be
\label{eq:dimensional}
\tau_{\rm diss} \sim \frac{2 \rho R^2}{5 \epsilon_B \epsilon_\Omega^2 \zeta}  \, ,
\ee
where the integral in Eq. (\ref{eq:def:taudiss}) is substituted by a volume-averaged bulk viscosity times a volume-averaged rotational 
deformation $\epsilon_\Omega  \sim \Omega^2 \sim \delta \rho/\rho$, and the NS moment of inertia as that of a uniform density sphere. 

We note some implications of Eqs. \ref{eq:dimensional}: {\it i)} $\tau_{\rm diss}  \sim \zeta^{-1}$: larger(smaller) values of the bulk viscosity coefficient imply a shorter(longer) dissipation time. Note that, since $\zeta \sim T^6$, the tilt angle evolution is very sensitive to the NS cooling history; {\it ii)} $\tau_{\rm diss} \sim \epsilon_B$, since $\zeta \sim \epsilon_B^{-2}$: the dissipation time is thus longer for larger ellipticities; 
{\it iii)} $\tau_{\rm diss} \gg T_p \approx {\rm P}_{\rm ms} / (\epsilon_{B, -3} \cos \chi)$ s: damping takes a large number of cycles, unless $\chi \approx \pi/2$.
\subsection{Tilt angle growth time}
\label{sec:growth-time}
The growth time of the tilt angle, $\tau_\chi$, is defined as
\be
\label{eq:tauchi}
\tau_\chi = \displaystyle \frac{{\rm sin} \chi} {\displaystyle \frac{d}{dt}{\rm sin}\chi}  = \frac{{\rm sin} \chi}{\dot{\chi} {\rm cos} \chi}\, ~
\ee
from which, using the expression for E$_{\rm pre}$ and its time derivative, we obtain (Dall'Osso et al. 2009)
\begin{equation}
\label{eq:tauchi-numbers}
\tau_\chi =  \displaystyle \frac{{\rm sin}^2 \chi}{{\rm cos}^2 \chi} \tau_{\rm diss} \, .
\ee
Combining these two expressions, the evolution equation for the tilt angle is readily obtained. 

\subsection{ Tilt angle evolution}
\label{sec:tilt-evolution}
For each of the three EoS's of Sec. \ref{sec:symmetry}, and assuming a $npe \mu$ composition, we calculated the integral in Eq. \ref{eq:def:taudiss} numerically, as described in detail by Dall'Osso \& Perna (2017), {\it i.e.} following the change of $\Delta \rho$ described in Sec. \ref{sec:energy-diss} as the NS switches from the low-frequency to the high-frequency regime of bulk viscosity. For millisecond spin periods and ellipticities $\epsilon_B \sim 10^{-3}$, as expected in newly born magnetars, bulk viscosity enters the high-frequency regime at temperatures $\lesssim 10^{10}$ K. The resulting expressions for $\tau_{\rm diss}$ have been used to calculate the temporal evolution of $\chi$. 

\subsubsection{Coupled tilt angle and spin evolution} Since $\tau_{\rm diss}$ depends on the precession frequency, hence on the NS spin, the evolution equations for $\chi$ and $\Omega$ are formally coupled. Dall'Osso et al. (2009) solved the equation for $\chi(t)$ under the assumption of a constant $\Omega$, {\i.e.} that spin-flip was much faster than the initial spindown due to the magnetic dipole. Here we extend their treatment, by adopting the newly derived expression for $\tau_{\rm diss}$ and solving the coupled evolution equations for $\chi(t)$ and $\Omega(t)$, without restrictions on the relation between the spin-flip and magnetic dipole spindown timescales.  
 \begin{figure}
\begin{center}
\includegraphics[scale=1.02]{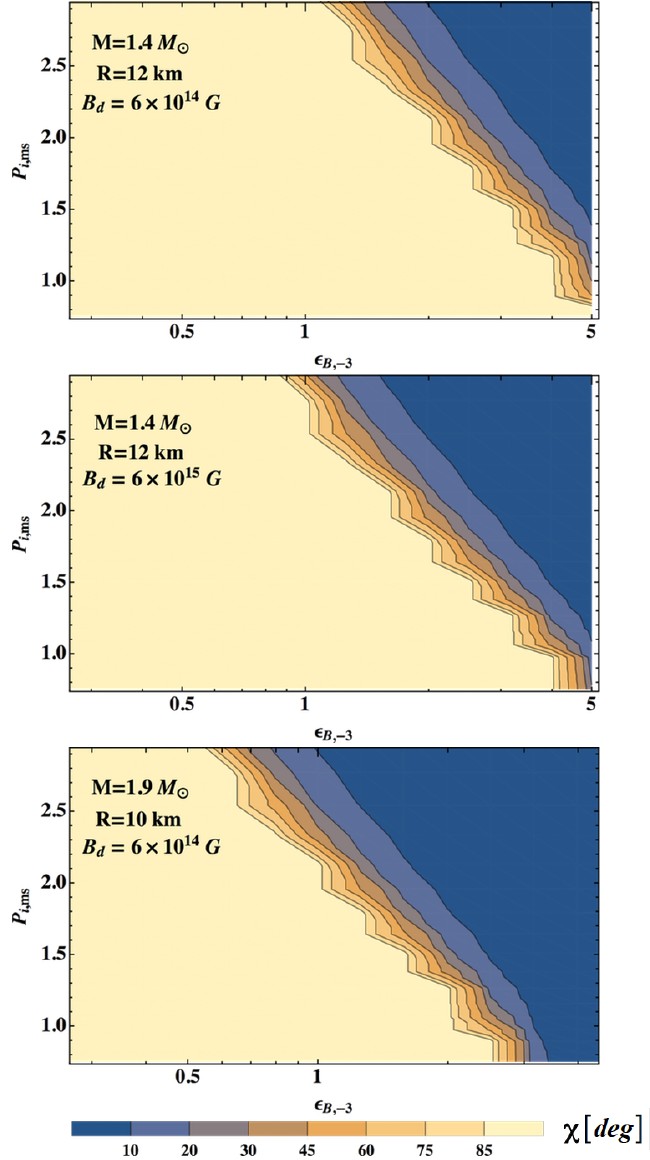}
\caption{Tilt angle $\chi$, calculated at $t \gtrsim 10^3$ s, as a function of the initial spin and magnetically-induced ellipticity, for EoS $II$ and three different NS parameter combinations (shown in the plots). The  parameter space is divided in two by a relatively narrow strip: most magnetars end up either  as almost orthogonal (yellow area, left) or almost aligned (blue area, right) rotators. A larger magnetic dipole ({\it middle panel}) has a minor impact demonstrating that spin-flip, when it occurs, is faster than the EM spindown. A larger mass, on the other hand, effectively delays spin-flip, slightly reducing the range in parameter space occupied by large tilt angles ({\it lower panel}). In the upper panel, the relation $P_{\rm ms} < -3.33 \log_{10} \epsilon_{B, -3} + 3.23$ approximates the dividing line at $\chi \approx 60^\circ$. Similarly, for the middle and lower panels we obtain, respectively, $P_{\rm ms} < -3.1 \log_{10} \epsilon_{B, -3} + 2.85$ and $P_{\rm ms} < -3.23 \log_{10} \epsilon_{B, -3} + 2.27$. }
\label{fig:angles}
\end{center}
\end{figure}
 \subsubsection{Numerical solutions}
 \label{sec:spindown-torques}
 We obtain the evolution equation for the spin frequency by considering that the NS spin energy is extracted by two mechanisms : {\it i)} magnetic dipole radiation - which acts right after the NS is born - at the rate (e.g., Spitkovski 2006)
\be 
\label{eq:spitkovski}
\dot{{\rm E}}_{\rm EM} =  - \frac{\mu^2 \Omega^4}{c^3} \left(1+{\rm sin}^2 \chi\right) \, ,
\ee
where $\mu = {\rm B}_{\rm d} R^3/2$ is the NS magnetic moment, B$_{\rm d}$ the dipole field strength at the magnetic pole and $\chi$ the tilt angle of the dipole field to the spin axis (assumed to be equal to the tilt of the axis of the interior toroidal field);  {\it ii)} GW emission,  at the rate (Jones \& Andersson 2001) 
\be
\label{eq:GWspindown}
\dot{{\rm E}}_{\rm GW} = - \frac{2}{5} \frac{{\rm G}}{c^5}  \left(I \epsilon_B \right)^2 \Omega^6 {\rm sin}^2 \chi \left(1+15 \sin^2 \chi\right) \, ;
\ee
we set $\sin^2 \chi \left(1+15 \sin^2 \chi\right)  \equiv \hat{F}(\chi)$ for later use.

The differential equation for $\Omega$ will thus be 
\be
\label{eq:omegadot}
\dot{\Omega} = \left(\dot{{\rm E}}_{\rm GW} + \dot{{\rm E}}_{\rm EM}\right)/I\Omega  .
\ee
~\\

\noindent
{\bf Tilt angle distribution at $t=10^3$ s --}
 By combining  Eqs. \ref{eq:omegadot} and \ref{eq:tauchi-numbers}, and using Eq. \ref{eq:def:taudiss} for $\tau_{\rm diss}$, we can calculate the numerical solutions for $\chi(t)$ and $\Omega(t)$ given the NS mass, radius, EoS (hence $\zeta$) and initial conditions ($\chi_i$, P$_{\rm ms, i}$, $\epsilon_B$, B$_{\rm d}$). Fig. \ref{fig:angles} illustrates results for EoS $II$, with the value of $\chi$ at $t=10^3$ s reported as a function of the initial spin period and NS ellipticity, in three representative cases: {\it a)} in the top panel, a 1.4 M$_\odot$ NS with a 12 km radius and B$_{\rm d} = 6 \times 10^{14}$ G are used; {\it b)} in the middle panel, the same mass and radius are considered, along with a much stronger dipole B$_{\rm d}= 5 \times 10^{15}$ G; {\it c)} the bottom panel shows results for an $1.9$ M$_\odot$ NS, with a 10 km radius and B$_{\rm d} = 6 \times 10^{14}$ G. Over most of the parameter space, the growth time of the tilt angle is $< 20$ s, and in no case it exceeds a few hundred seconds. 

In all panels, two regions are apparent: almost orthogonal or nearly aligned rotators, separated by a relatively narrow strip of intermediate cases. The upper and middle panels show clearly that the separation is almost insensitive to the strength of the magnetic dipole field, 
confirming that spin-flip is, in general, faster than magnetic dipole spindown (Dall'Osso et al. 2009). The upper and lower panels, on the other hand, show a small, yet noticeable influence of the NS compactness: the more massive and smaller NS tends to dissipate more slowly, resulting in a slightly larger region of aligned rotators in parameter space (see caption for more details).  
Spin-flip fails in the high-$\epsilon_B$ and long P$_{\rm ms}$ sector of parameter space, owing to the freezing of viscous dissipation when the dissipation time is longer than the cooling time. At birth, both timescales  scale as $T^{-6}$, so whether this condition is realized or not is determined by initial values of $\epsilon_B$ and P$_{\rm ms}$. Once the switch to the high-frequency regime of bulk viscosity has completed, $\tau_{\rm diss}$ becomes even more sensitive to $T$, progressively freezing  the value of the tilt angle. Since, in this regime, the dissipation time is $\sim \epsilon_B$, large-ellipticity NSs are those mostly affected by the freezing.

\noindent
The density-averaged bulk viscosity coefficient for EoS $II$ is $\approx 3$ times larger than the value in Eq. \ref{eq:zeta-stand}. For EoS $I$ it is almost twice as strong, making an even larger region of parameter space accessible to orthogonal rotators. For EoS $III$, it is just 2/3 of the one adopted in Fig. \ref{fig:angles}, leading to  a small reduction of the interesting parameter range.

\section{GW \& EM transients}
\label{sec:transients}
The NS spin energy is E$_{\rm spin} = (1/2) I \Omega^2$. The moment of inertia can be well approximated by the polynomial\footnote{Valid for $\beta >0.1$ and maximum NS mass $\ge 1.97$ M$_{\odot}$.} (Lattimer \& Prakash 2016)
\be
\label{eq:inertia}
I \approx M R^2 \left(0.247+0.642 \beta + 0.466 \beta^2\right) \, ,
\ee
as a function of the compactness $\beta = G M/(c^2 R)$. Because the maximum NS spin frequency is also expressed by a nearly universal relation (Lattimer \& Prakash 2016),
\be
\label{eq:numax}
\nu_{\rm max} \approx 1.08~{\rm kHz}~\left(\frac{M}{1.4M_\odot}\right)^{1/2} \left(\frac{R}{10{\rm km}}\right)^{-3/2} \, , 
\ee
where $M$ and $R$ refer to the non-rotating configuration, then the maximum spin energy of a NS can be expressed, in terms of (\ref{eq:inertia}) and (\ref{eq:numax}), as
\be
\label{eq:maxenergy}
\begin{split}
{\rm E}_{\rm spin, max} =4.6 \times 10^{52} \left(\frac{M}{M_\odot}\right)^2 \left(\frac{R_*}{10~{\rm km}}\right)^{-1} \\
\left(0.247 + 0.642 \beta + 0.466 \beta^2\right) ~{\rm erg} \, . 
\end{split}
\ee
The maximum NS mass is $\gtrsim 2$ M$_\odot$ (Antoniadis et al. 2013): thus, E$_{\rm spin, max}$ can range from $\lesssim 3 \times 10^{52}$ erg, for a M = 1.4 M$_\odot$, R$=12$ km NS, to $\sim 10^{53}$ erg in extreme cases.

The two torques described in Sec. \ref{sec:spindown-torques} draw spin energy and channel it into the EM and GW windows, respectively.  
Eq. \ref{eq:spitkovski} gives the initial spindown time due to magnetic dipole emission, $\tau_{\rm em} \sim 1.7~{\rm day}~P^2_{i, {\rm ms}} B^{-2}_{{\rm d}, 14} \left(1+\sin^2 \chi\right)^{-1}$, $P_i$ being the birth spin period. From Eq. \ref{eq:GWspindown} we get the GW-driven spindown time,  $\tau_{\rm GW} \sim 3.3~{\rm day}~P^4_{i, {\rm ms}} \epsilon_B^2 \hat{F}(\chi)^{-1}$. Given  the large spin energy reservoir and short timescales involved, bright EM and/or GW transients may occur when a highly magnetised, millisecond spinning NS is formed.  The initial conditions determine the relative strengths of the EM and GW spindown luminosities. 

The solutions for $\chi(t)$ and $\Omega(t)$ from Sec. \ref{sec:spindown-torques} are needed in order to calculate spindown luminosities as a function of time in both the GW and EM window (Eqs. \ref{eq:spitkovski} and \ref{eq:GWspindown}); in this way, it is possible characterise both types of transients and determine  their detectability.
\subsection{GW transient signals}
\label{sec:GW-signals}

Our numerical solutions for $\chi(t)$ extend the analytical results of Dall'Osso et al. (2009). On the one hand they confirm that, for a large portion of  parameter space, efficient GW emission is favored by the tilt angle quickly attaining large values ($> 60^\circ$), before the NS spin energy can be drained by EM torques.  
On the other hand, they reveal that bulk viscous dissipation will drop faster than previously calculated, limiting the growth of the tilt angle at large $\epsilon_B$, despite the larger bulk viscosity coefficient calculated here. In particular, if\footnote{The exact value depending also on the NS spin.} $\epsilon_{B, -3} \gtrsim 5 $ the tilt angle {remains small} and, accordingly, the GW emission efficiency has a sharp drop.

To update earlier results on the strength and detectability of the expected GW signals, we first estimate the signal-to-noise ratio (S/N) for a one-detector ideal match-filtered search (e.g. Owen \& Lindblom 1998). The orientation- and position-averaged strain is (Finn \& Chernoff 1993)
\be
\label{eq:h}
h^2_a(f) = \frac{2 \pi^4 G^2 I^2 \epsilon^2_B}{5 c^8 D^2} f^4 \hat{F}(\chi) 
\ee
where $f = \Omega/\pi$ is the GW signal frequency and $D$ the source distance. The signal-to-noise ratio is thus 
\be
\label{eq:signal}
S/N = 2 \sqrt{\int \frac{\tilde{h}^2_a(f)}{S_h(f)}~df } \, .
\ee
In Eq. (\ref{eq:signal}), $\tilde{h}(f)$ is the Fourier transform of the instantaneous strain, and $S_h(f)$ the one-sided noise spectral density of the detector. For Advanced LIGO/Virgo 
we adopted the design sensitivity curve\footnote{Data files from https://dcc.ligo.org/LIGO-P1200087-v42.} of Abbott et al. (2017b).
~\\

\noindent
{\bf Detection by Advanced LIGO/Virgo -- } 
We calculated Eq. \ref{eq:signal} in the case of EoS $II$, for two values of the NS magnetic moment, two combinations of the NS mass and radius, and adopting a standard distance of 20 Mpc. Results are shown in Fig. \ref{fig:SN-Virgo} (see caption for details). The GW signals last from $\sim$ a few hours up to $\gtrsim$ a day (depending on the  values of P$_{\rm ms}$, $\epsilon_B$ and B$_{\rm d}$), during which the frequency decreases by a factor $\sim 2-3$. $S/N$-values in Fig. \ref{fig:SN-Virgo} were calculated by integrating the signals for the first 12 hrs after the NS is born. We verified that, for the weaker signals, increasing the time span up to $\sim 24$ hrs can result in a $\sim$ 25\% gain in S/N, while for the most powerful events the signal is concentrated in less than 12 hrs (our results are thus somewhat conservative for the lower-S/N signals). 

\noindent
Fig. \ref{fig:SN-Virgo} shows that a stronger B$_{\rm d}$ reduces the detectable region in parameter space and shifts it towards larger $\epsilon_B$, as it increases the EM over the GW torque. Increasing the NS mass has a comparable effect, but shifts the detectable region to lower ellipticities. This results from the combination of a larger moment of inertia,  which slightly favors the GW over the EM torque, and of incomplete flipping (see Fig. \ref{fig:angles}), which reduces the GW efficiency at larger $\epsilon_B$. 
\begin{figure}
\begin{center}
\includegraphics[scale=0.33]{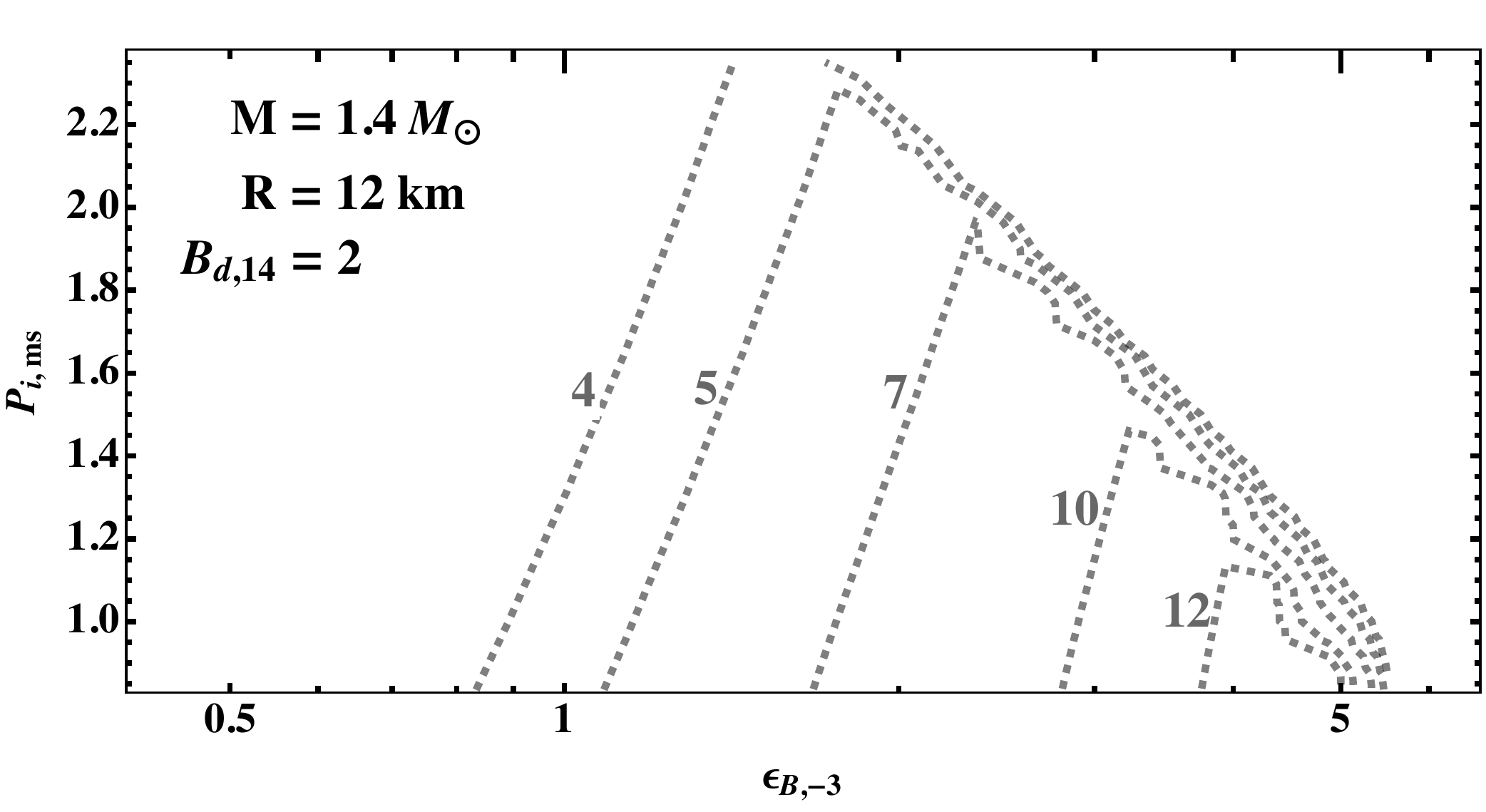}
\includegraphics[scale=0.33]{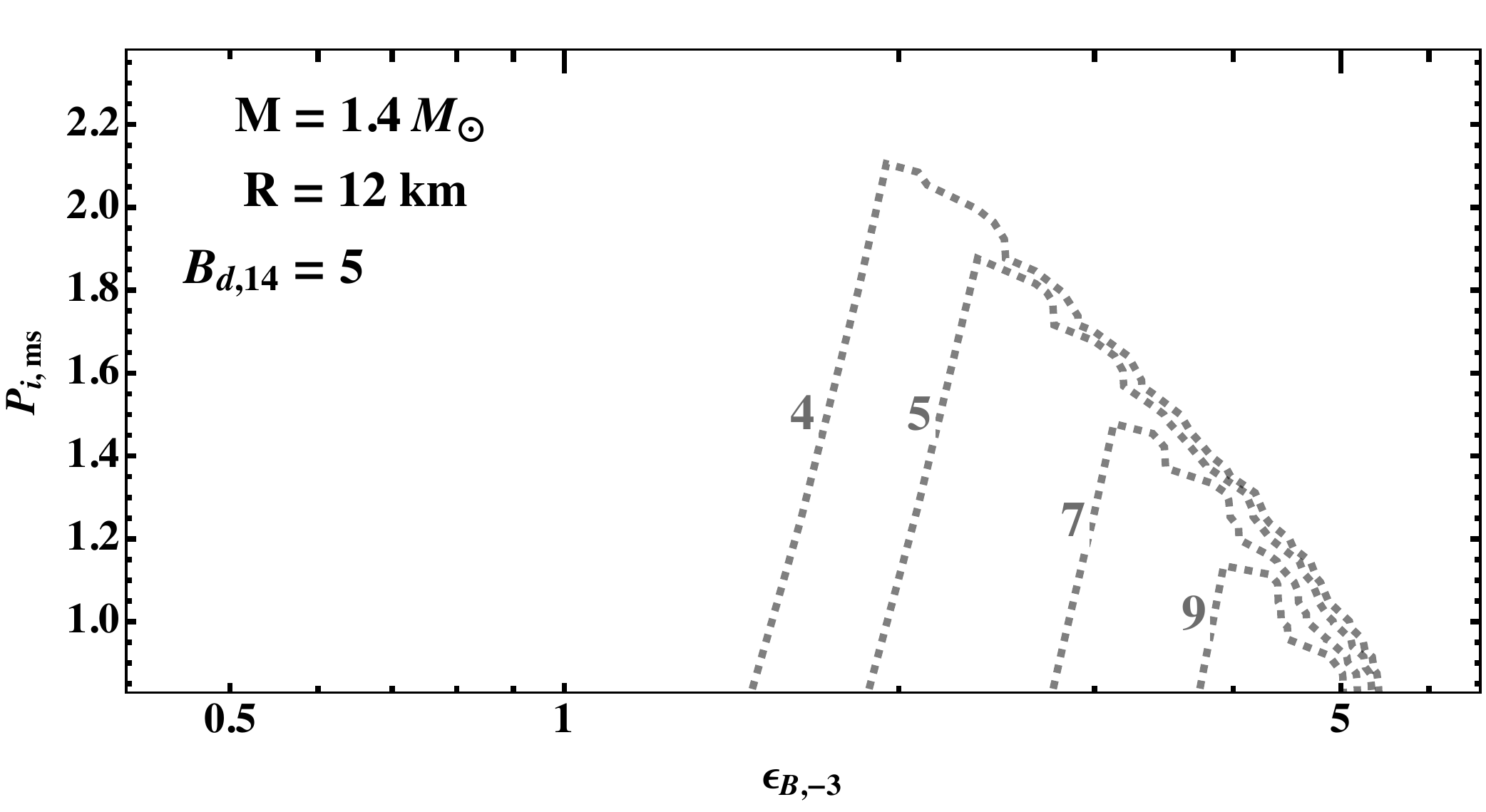}
\includegraphics[scale=0.33]{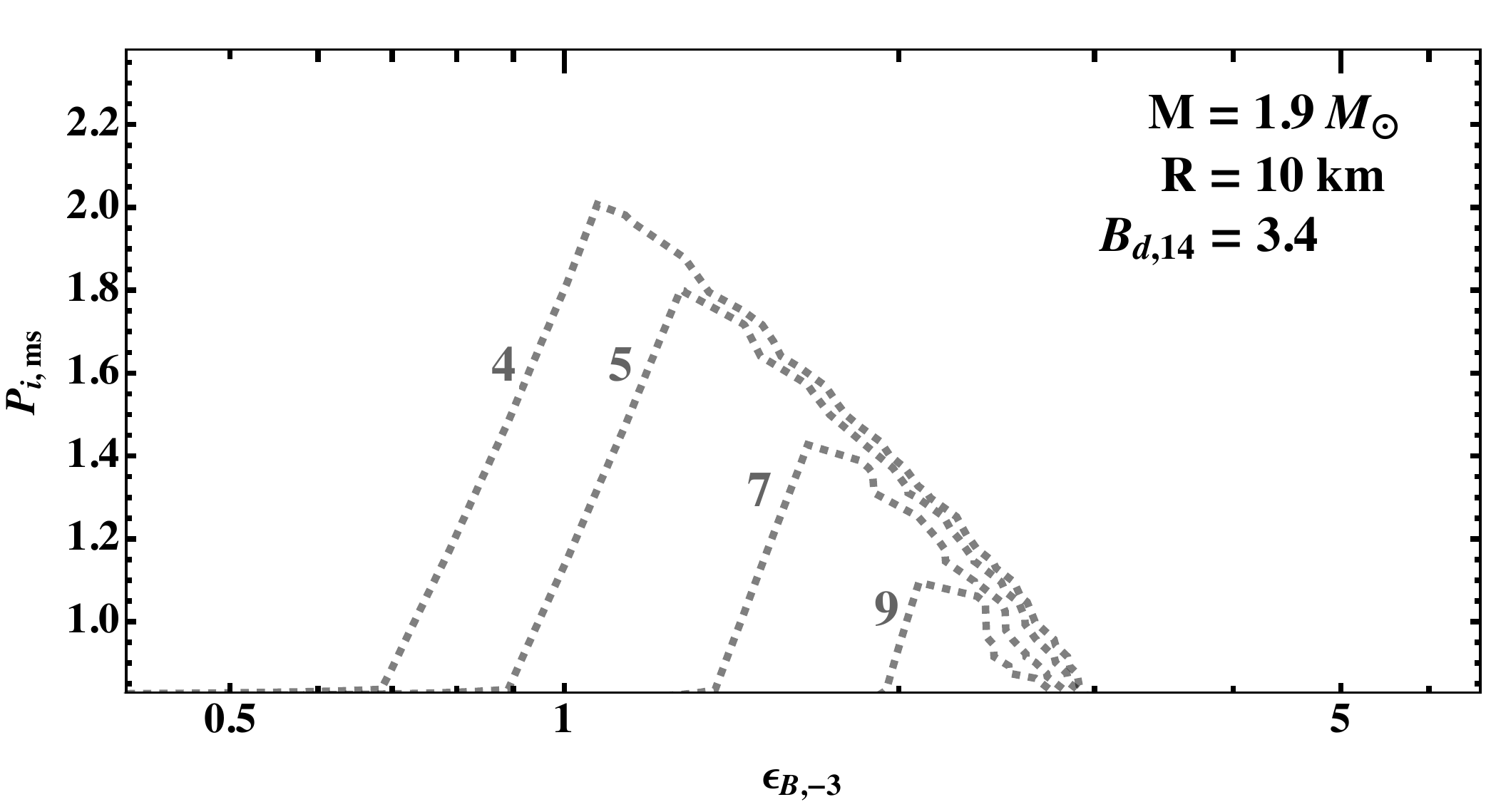}
\caption{The orientation- and position-averaged S/N of a newborn magnetar at 20 Mpc, for EoS $II$ and a single-detector matched-filter search, as a function of P$_{\rm ms}$ and $\epsilon_B$. Signals are integrated for $\Delta t=12$ hrs. We adopted two different values of the magnetic dipole moment and two choices of the NS mass and radius (see plots). In the upper and lower panels, the same value of $\mu_{32}=1.7$  is used: different values of $B_{\rm d}$ reflect different NS radii. The positive slope side of the contours reflects magnetars whose tilt angle becomes quickly $\sim 90^\circ$: they efficiently emit GWs with a strain $\propto \epsilon_B$. The negative slope side corresponds to sources in which cooling freezes spin-flip before the tilt angle becomes large: they emit GWs with lower efficiency, which drops at higher $\epsilon_B$.}
\label{fig:SN-Virgo}
\end{center}
\end{figure}
Finally, for a fixed value of B$_{\rm d}$, the maximum distance at which S/N is above a given threshold scales almost linearly with $1/P_{\rm i, ms}$ and $\epsilon_B$. A sharp cut off in the maximum distance occurs for $\epsilon_B$-values beyond the descending branch of the S/N curves in Fig. \ref{fig:SN-Virgo}.

The long duration and strong spin-down of these {\it time-reversed} ``chirps" pose new technical challenges, not fully addressed by current detection algorithms. While a complete discussion of these problems is clearly beyond the scope of this work, we summarize them here, in order to assess realistic perspectives for signal detection. 

\noindent
Even with match-filtering, actual signal searches with unknown phase parameters have an ${\cal F}$-statistics maximum S/N that is a factor $\sqrt{2}$ lower than the optimal value of Eq. \ref{eq:signal} (Eqs. 31, 64, 112 in Jaranowski \& Kr\'{o}lak 2000). In reality, the search for these GW signals will be limited by available computing power, and will have to be carried out by using sub-optimal methods, leading to a further loss of sensitivity (e.g., Thrane et al. 2011, Prix et al. 2012, Coyne et al. 2016). 
In a semi-coherent search, short data sets can be analyzed coherently and then combined incoherently to increase sensitivity. 
However, this is less sensitive than a coherent search by a factor $\sim N^{-1/4}$, where $N = T_{\rm obs}/T_{\rm short}$ is the number of short data sets in which the whole observation is split. 
The potential of these methods can be enhanced in an hierarchical scheme, in which candidate events are followed up with increasingly selective criteria and a finer tiling in a smaller region of parameter space. As a specific example, we assumed a search done with the frequency Hough-transform, which extends the existing calculations for continuous waves (Astone et al. 2014) to periodic signals slowing down on a $\sim 10^4-10^5$ s timescale (Miller et al., in preparation). We adopted $\epsilon_{B, -3} \sim 1$, $I_{45} = 1.4$, an initial spin period of 1 ms, $T_{\rm shor} \lesssim$ 100 s and standard values for (i) the threshold on the critical ratio for candidate selection on the Hough map (CR$_{\rm thr} \sim 5$), which sets the false alarm probability; (ii) the threshold for peak selection on the equalized power spectra $\theta_{\rm thr} =2.5$ (see Astone et al. 2014). The loss in sensitivity translates into a factor $\sim$ 5-6 smaller range with respect to the value adopted in Fig. \ref{fig:SN-Virgo} (Miller et al. in preparation).
~\\

\noindent
{\bf Detection by the ET --}  The sensitivity of third generation detectors, like the Einstein Telescope (ET), will improve significantly over that of Advanced LIGO/Virgo. In the ET-D configuration, for which $S_0 \approx 3.6 \times 10^{-49}$ Hz$^{-1}$ at 1 kHz and the 
a sensitivity gain by a factor $\sim 8$ can be anticipated. Therefore, the curves of Fig. \ref{fig:SN-Virgo} would hold for a distance of $\sim 160$ Mpc, whereas taking into account the sensitivity loss of semi-coherent searches, the corresponding range would become $\sim 25-30$ Mpc. 
~\\

\noindent
{\bf Expected event rate --}  
Millisecond spinning magnetars may be formed either in the core-collapse (CC) of massive stars or in binary NS mergers. Li et al. (2011) estimate a rate $\sim 0.7 \times 10^{-4}$ yr$^{-1}$ Mpc$^{-3}$ for all CC SNe in the local universe. To average out local over-densities, we integrate the rate over 60 Mpc, within which $\sim$ 65 CC SNe per year are expected. Using the cumulative blue light distribution in the local universe (Kopparapu et al. 2008; Abadie et al. 2010) as a proxy for the star formation rate, the above translates to a magnetar birth rate of (i) $\gtrsim$  0.3 yr$^{-1}$ within 20 Mpc, if they represent $\gtrsim$ 10\% of all NSs formed in CC SNe (note that Stella et al. 2005 estimated $\lesssim$ 1 yr$^{-1}$, based on the energetics of magnetar Giant Flares); (ii) $\gtrsim$ 0.01 yr$^{-1}$ within 4 Mpc, the sub-optimal horizon for Advanced LIGO/Virgo; (iii)
$\sim (0.5-1)$ yr$^{-1}$ within 25-30 Mpc, the estimated range of third generation detectors. 

Newborn magnetars formed in BNS mergers  could either be stable objects, or supra/hyper-massive NS bound to collapse to black holes after loosing some of their centrifugal support (e.g. Giacomazzo \& Perna 2013, Metzger 2017). In the latter case, their GW signals, albeit shorter lived, may be especially rich of information about the NS EoS (Dall'Osso et al. 2015; Piro et al. 2017). In the former case, the signal would be slightly stronger than calculated here - because the NS is close to the maximum mass - and would thus allow for a somewhat larger horizon, $\lesssim 40$ Mpc with third generation detectors (adopting the numbers from Dall'Osso et al. 2015, and factoring in the sensitivity loss of realistic searches, as discussed above).  The event rate for BNS mergers is estimated to be $\sim (320-4700)$ Gpc$^{-3}$ yr$^{-1}$ (Chruslinska et al. 2017 and references therein), implying a rate $\sim (0.9 - 13) \times 10^{-4}$ yr$^{-1}$ within the sub-optimal horizon for Advanced LIGO/Virgo, and $\sim (0.09-1.3)$ yr$^{-1}$ within the sub-optimal horizon for third generation detectors (see e.g. Dall'Osso et al. 2015, Piro et al. 2017 for a discussion of the fraction of BNS mergers that may produce stable/supra-massive NSs). In the latter case, the stronger ``chirp" from the inspiral would serve as a trigger for a targeted search of the signal emitted by the newly formed NS.

\subsection{EM transients}
\label{sec:EM-transients}
The GW signals from magnetars born in CC are expected to be associated to SN explosions (e.g. Thompson \& Duncan 1993, Gaensler et al. 1999). Within the ranges estimated in Sec. \ref{sec:GW-signals} for Advanced and third generation interferometers, these SNe would be easily identified in the optical/NIR. This makes prospects for multi-messenger studies of such GW events especially promising. 
For magnetars originating from binary NS mergers, the same two types of EM counterparts predicted for the merger are expected: a prompt, short gamma-ray burst (GRB) for favorable viewing angles (e.g. Eichler et al. 1989) and/or a kilonova on a timescale $\sim$ day, for a wide range of viewing angles (e.g., Li \& Paczy\'{n}ski 1998, Metzger 2017; Coulter et al. 2017,  Arcavi et al. 2017, Cowperthwaite et al. 2017, Kasen et al. 2017).

\noindent
The NS spin energy may contribute powering the SN or other phenomena, if magnetic dipole spindown is dominant, as extensively discussed in the literature. Magnetars formed in core-collapse may lead to the production of a long-GRB (e.g. Thompson et al. 2004, Bucciantini et al. 2006, Metzger et al. 2007, 2011), a shallow decay phase in the GRB early afterglow (e.g. Zhang \& Meszaros 2001, Dall'Osso et al. 2011, Bernardini et al. 2012) or a Super-Luminous Supernova (SLSN; Kasen \& Bildsten 2010, Greiner et al. 2015). The spin energy of a magnetar formed in a binary NS merger may power the extended emission following (some) short GRBs (Metzger et al. 2008, 2011). 
The co-existence of a GW torque has not been considered yet in any of such scenarios; however, GW-driven spindown was considered by, e.g. Dall'Osso \& Stella 2007 and Dall'Osso et al. 2009, in relation to the energetics of SN remnants associated to galactic magnetars. We plan to address this issue in greater detail in a future study.
\section{Conclusions}
\label{sec:conclusions}
We improved on previous work on the 
role of the spin-flip instability in the GW emission from newborn magnetars, by: a)  calculating the coefficient of bulk viscosity, the dissipative process driving the instability, for $npe \mu$ matter, with various realistic EoS's; b) introducing a prescription for the way in which fluid compressibility drops as the cooling NS switches between the low- and high-frequency limits of bulk viscosity; c) deriving the first self-consistent solution of the coupled evolution equations for the NS tilt angle (``spin-flip") and spin frequency, under the effect of bulk viscous dissipation, and of GW and EM spindown torques. Based on that, we calculated the detectability of the GW signal with Advanced LIGO/Virgo and future third generation detectors, as a function of the magnetically-induced ellipticity and initial spin period of the NS.

\noindent 
Our main conclusions are: i)
the bulk viscosity coefficient of NS matter with a realistic EoS and chemical composition is, in general, larger than the standard expression valid for pure $npe$ matter. This makes bulk viscous dissipation more efficient than previously calculated, and the ``spin-flip" instability accordingly faster; ii) the ``spin-flip" instability freezes, and the tilt angle stops growing, as the NS cools below $\sim (8-9) \times 10^9$ K, due to the decreasing compressibility of the $npe \mu$ fluid. iii) at spin periods $\lesssim$ 2 ms, spin-flip will cause a fast growth of the tilt angle $\chi$, in turn causing strong GW emission. At large ellipticities, on the other hand, the tilt angle growth time is proportionally longer, and spin-flip freezes before $\chi$ has evolved significantly. We find that, for $\epsilon_{B, -3} \gtrsim 5$, GW emission is quenched because the tilt angle remains close to its (small) initial value;
iv) in realistic data analyses with sub-optimal sensitivity, Advanced LIGO/Virgo-class detectors can capture the GW signal of a millisecond spinning, magnetically-distorted NS up to a distance $\sim (3-4)$ Mpc $\epsilon_{B,-3}/{\rm P}_{\rm ms}$, 
provided that the dipole B-field is B$_{\rm d} \lesssim 3 \times 10^{14}$ G. Magnetars are expected to form in $\gtrsim$ 10 \% of CC SNe, implying one event per $\sim$ 30-100 yrs within that distance range. 
v) Third-generation interferometers, with a sensitivity improved by a factor $\sim$ 8 in the relevant frequency range, will push the horizon of even sub-optimal searches to $> 30$ Mpc, within which an event rate $\gtrsim (0.5-1)$  yr$^{-1}$ can be expected.
In addition, this expanded horizon would likely include an interesting number of BNS mergers ($\sim 0.09-1.3$) yr$^{-1}$ according to current best estimates), increasing the chances of detecting magnetars formed in binary NS mergers, the exact rate of which depends on the uncertain fraction of mergers that can produce a stable, or long-lived, NS.  

For magnetars formed in the core-collapse of massive stars the accompanying optical/NIR SN should be easily detectable within the horizon of Advanced and third generation interferometers, leading to a robust association between the GW signal and the EM counterpart. For magnetars formed in BNS mergers, an optical/NIR kilonova emission should be expected for most viewing geometries, along with a (short) $\gamma$-ray burst for a favourable viewing angle and/or for sufficiently small distances, as the event GW170817 has demonstrated (e.g. Abbott et al. 2017e; Coulter et al. 2017, Arcavi et al. 2017; Alexander et al. 2018). 
The spin energy of a millisecond NS can provide additional power to the EM emission of these events, in a manner similar to that discussed for e.g. long-GRBs and SLSNe. 
In all cases, detection of the characteristic GW {\it time-reversed} ``chirp" associated to a newborn magnetar would give an unambiguous confirmation of the nature of the central engine. The development of much-needed search algorithms and strategies for the detection of such signals is urged.

\newpage

{\bf Acknowledgments}.
SD acknowledges support from NSF award AST-1616157. LS acknowledges financial contribution from agreement ASI-INAF I/037/12/0.

\clearpage

\end{document}